\documentclass[preprint,eqsecnum,aps,prd,nofootinbib,showpacs]{revtex4}

\begin{document}

\tolerance=5000

\newcommand{\mat}[2][rrrrrrrrrrrrrrrrrrrrrrrrrrrrrrr]{\left[
\begin{array}{#1}
#2\\
\end{array}
\right]}

\def\be{\begin{equation}}
\def\ee{\end{equation}}
\def\bea{\begin{eqnarray}}
\def\eea{\end{eqnarray}}

\title{The non-zero energy of 2+1 Minkowski space}
\author{Donald Marolf}
\email{marolf@physics.ucsb.edu}
\author{Leonardo Pati\~no}
\email{leo_p@physics.ucsb.edu}
\affiliation
   {Physics Department, \\
    UCSB, \\
    Santa Barbara, CA 93106}
\date{\today}

\begin{abstract}
We compute the energy of 2+1 Minkowski space from a covariant action principle.   
Using Ashtekar and Varadarajan's characterization of 2+1 asymptotic flatness, we first show that  the 2+1 Einstein-Hilbert action with Gibbons-Hawking boundary term is both finite on-shell (apart from past and future boundary terms) and stationary about solutions under {\it arbitrary} smooth asymptotically flat variations of the metric. Thus, this action provides a valid variational principle and no further boundary terms are required.  
We then obtain the gravitational Hamiltonian by direct computation from this action.  The result agrees with the Hamiltonian of Ashtekar and Varadarajan up to an overall addititve constant.  This constant is such that 2+1 Minkowski space is assigned the energy $E_{Mink_{2+1}} = -\frac{1}{4 G}$, while the upper bound on the energy becomes $E \le  0$.  Any variational principle with a boundary term built only from the extrinsic and intrinsic curvatures of the boundary is shown to lead to the same result.  Interestingly, our result is {\it not} the $\Lambda \rightarrow 0$ limit of the corresponding energy  $E_{AdS_{2+1}} = -\frac{1}{8G}$ of  2+1 anti-de Sitter space. 
\end{abstract}

\maketitle

\section{Introduction}
\label{intro}

\vspace{1cm}

In covariant approaches to quantum mechanics, classical mechanics is recovered
through a semi-classical approximation in which the path
integral is dominated by stationary points of the action.   Here it is critical that the action 
be stationary under {\it all} variations
tangent to the space of paths over which the integral is
performed. Thus one must consider all variations which
preserve any boundary conditions and not just, for example, variations of
compact support. In particular, requiring the action to be
stationary should yield precisely the classical equations of
motion, with all boundary terms in the associated computation
vanishing on any allowed variation.

We are concerned here with gravitational systems.  The most familiar action for gravity is the Einstein-Hilbert action with Gibbons-Hawking boundary term,
\be
S_{EH+GH}={1 \over 16\pi G}\int_{\cal{M}}\sqrt{-g}R +{1 \over 8\pi G}\int_{\partial \cal{M}}\sqrt{-h}K, \label{S}
\ee
where $h_{ij}$ is the induced metric on a timelike boundary $\partial \cal{M}$ and $K_{ij}$ is the extrinsic curvature of this boundary, with trace $K=K_{ij}h^{ij}$.
 However, for spacetimes with non-compact spatial slices the action (\ref{S}) is generally not stationary under the full class of allowed variations about solutions.  In the asymptotically AdS context, it was shown in \cite{PK} that adding the AdS counter-terms of \cite{skenderis,Balasubramanian:1999re,KS2,KS3,KS4,KS5,KS6,KS7,KS8}  results in a fully stationary action.  In addition, 
\cite{Mann:2005yr} showed that a fully stationary action for asymptotically flat spacetimes of dimension $d \ge 3+1$ can be defined by adding an appropriate additional boundary term
 to (\ref{S}).  In the asymptotically flat cases most closely related to what we study below, this additional boundary term reduces to the term proposed in \cite{Mann:1999pc}  (see also \cite{Kraus:1999di})\footnote{Fully stationary variational principles were developed somewhat earlier in other formalisms.  For example, the Regge-Teitelboim construction \cite{RT} defines a fully stationary Hamiltonian $H$, which immediately leads to a fully stationary variational principle of the form $S = \int dt [\left( \int_\sigma \tilde \pi^{ab} \dot{g}_{ab} \right) -H]$.   This method was applied to asymptotically AdS spaces in \cite{HT,BH}.  Similarly, in 3+1 dimensions \cite{eeA,ABL} provided a fully stationary variational principle in the Palatini formalism and further showed that the 3+1 decomposition of that action led to the analogue of the ADM Hamiltonians.}.

Here we address the case of 2+1 asymptotically flat dimensions.  In contrast to the higher dimensional case, we find that this Gibbons-Hawking term suffices and that no further boundary terms are required to make the action fully stationary.  In particular, 
following the characterization of 2+1 asymptotic flatness given in \cite{Ashtekar:1993ds} (reviewed in section \ref{review} below), we show in section
\ref{action} that (\ref{S}) is both i) finite  on the space of solutions (apart from past and future boundary terms) and ii) stationary on solutions under any variation which preserves this definition of asymptotic flatness.  

Thus, the  action (\ref{S})  provides a valid variational principle.  As a result, one may calculate a gravitational Hamiltonian as the Legendre transform of the corresponding Lagrangian.  The calculation is essentially that of \cite{Hawking:1995fd}, but  without a background subtraction term.  We perform this calculation in section \ref{Hamiltonian}.  The result agrees with the Hamiltonian of Ashtekar and Varadarajan \cite{Ashtekar:1993ds}  up to an overall addititve constant.  This constant is such that our definition assigns 2+1 Minkowski space the non-zero energy $E_{Mink} = -\frac{1}{4\pi G}$, while the upper bound on the energy becomes $E \le  0$.  From the Hamiltonian form of the action it is straightforward to show (section \ref{other}) that the same result also follows from the boundary stress tensor associated with (\ref{S}).  We close with a brief discussion in section \ref{disc}.

\section{Asymptotic flatness for $d=2+1$}
\label{review}

We now review the definition of 2+1 asymptotic flatness given in \cite{Ashtekar:1993ds}.
Recall that the solution 
\be
ds^2=-dt^2+r^{-8GM}(dr^2+r^2d\theta^2) \ \ \ {\rm for} \ r > 0 \label{pp}
\ee
represents a point particle of mass $M$ at the origin \cite{Deser:1983tn}.
Here, G is the three dimensional Newton's constant and the coordinates  take values as follows: $t\in(-\infty, +\infty)$, $r\in [0,\infty)$, and $\theta\in [0,2\pi)$; i.e., the coordinates range over the usual values associated with familiar cylindrical coordinates.

Curvature does not propagate in $2+1$ dimensions, so for any solution the Riemann tensor vanishes wherever $T_{\mu\nu}=0$.  In particular, the metric (\ref{pp}) is flat everywhere except at the origin where the particle is located. To see this, we can perform the coordinate transformation
\be
\rho\equiv {r^\alpha\over\alpha} , \;\;\;\; \bar\theta\equiv\alpha\theta, \;\;\;\; {\mathrm{with}} \;\;\;\; \alpha\equiv1-4GM, \label{coot}
\ee
to cast the metric in the form
\be
ds^2=-dt^2+d\rho^2+\rho^2d\bar\theta^2, \label{ppfl}
\ee
from which the flatness of the metric is apparent.

Given that $\bar\theta \in [0,\alpha)$, one also sees that there is a deficit angle which, despite the local flatness of the solution away from the origin, makes this spacetime not globally equivalent to Minkowski space.  As a result, the general definition of $2+1$ asymptotic flatness is chosen to allow metrics which asymptotically approach such conical spacetimes.  In particular,  Ashtekar and Varadarajan  \cite{Ashtekar:1993ds} consider a 2+1 spacetime to be asymptotically flat  if the line element admits an expansion of the form
\be
ds^2=-(1+{\cal O}(1/r))dt^2+r^{-\beta} [(1+{\cal O}(1/r))dr^2+r^2(1+{\cal O}(1/r)) d\theta^2] +{\cal O}(r^{-1-\beta/2})dtd\theta , \label{afg}
\ee
for large positive $r$ if $\beta\in [0,2)$. We use this characterization of asymptotic flatness below.

\section{Finiteness and stationarity of the action on the space of solutions}
\label{action}

In this section, we show that the action (\ref{S}) is both i) finite apart from past and future boundary terms and ii) fully stationary on the space of asymptotically flat solutions, under any variation preserving asymptotic flatness.  We then compare the situation in 2+1 dimensions with that of higher dimensions, in which more complicated boundary terms are required.  We will see that the higher dimensional boundary terms do not extend to the 2+1 case in an obvious way.
  
Let us begin by stating precisely the prescription for computing the boundary term associated with the timelike part of the boundary $\partial {\cal M}$, which we call $\partial {\cal M}_\infty$.   The notation  $\partial {\cal M}_\infty$ is shorthand for some one-parameter family of boundaries of regions ${\cal M}_\Omega \subset {\cal M}$. We take the regions ${\cal M}_\Omega$ to be  an increasing family, so that they  satisfy ${\cal M}_\Omega \subset {\cal M}_{\Omega'}$ whenever $\Omega < \Omega'$, and such that they cover ${\cal M}$, i.e. $\cup_\Omega{\cal M}_\Omega = {\cal M}$. Any such family represents a way of introducing a cut-off for the spacetime ${\cal M}$ and then removing it as $\Omega \rightarrow \infty $. Here we  take $\Omega=r+{\cal O}(r^0)$. This is the analogue of what was termed a `cylindrical spatial cut-off' in \cite{Mann:2005yr}. The definition of $\Omega$ has been chosen so that the induced line element on any $\Omega=constant$ surface is of the form
\be
ds^2_{\partial {\cal M}}= h_{ij} dx^i dx^j = -(1+{\cal O}(1/r))dt^2+(r^{2-\beta}+{\cal O}(r^{-1-\beta}))d\theta^2 +{\cal O}(r^{-1-\beta/2})dtd\theta , \label{afh}
\ee
with inverse metric
\be
\label{invh}
h^{ij} = \mat{ h^{tt} & h^{t\theta} \\ h^{\theta t} & h^{\theta \theta} } = \mat{ -1 + {\cal O}(1/r)  \ \ & {\cal O}(r^{-3+\beta/2}) \\ {\cal O}(r^{-3+\beta/2}) \ \ & r^{\beta -2} + {\cal O}(r^{\beta-3})  }.
\ee

To obtain a finite action, we also require some cut-off on the integral over time.  Here we simply restrict the coordinate $t$ of (\ref{afg}) to some finite range $t \in [T_-,T_+]$.  This corresponds to what was called a `cylindrical temporal cut-off' in \cite{Mann:2005yr}.  

One might also attempt to consider an anologue of the hyperbolic temporal cut-off from \cite{Mann:2005yr}.  However, this is less natural in the 2+1 context due to the lack of asymptotic Lorentz invariance \cite{Ashtekar:1993ds,MH}.  Furthermore, hyperbolic cut-offs are less natural in the context of the Hamiltonian methods we wish to apply.  For these reasons we consider only the cylindrical cut-off below.

\subsection{The action is finite and stationary}

To fully define the space of asymptotically flat geometries, we need to specify the asymptotic behavior of the stress-energy tensor. Following  \cite{Ashtekar:1993ds}, we take the components of $T_{\mu\nu}$ in cartesian coordinates to be of order ${\cal O} (r^{\beta-3})$, so that in polar coordinates we have
\be
T_{\mu\nu} = \mat{T_{tt}&T_{tr}& T_{t\theta} \\  T_{rt}&T_{rr}& T_{r\theta} \\  T_{\theta t}&T_{\theta r}& T_{\theta\theta} } \sim \mat{ {\cal O}(r^{\beta-3}) & {\cal O}(r^{\beta-3}) & {\cal O}(r^{\beta-2}) \\
{\cal O}(r^{\beta-3}) & {\cal O}(r^{\beta-3}) & {\cal O}(r^{\beta-2}) \\ {\cal O}(r^{\beta-2}) & {\cal O}(r^{\beta-2}) & {\cal O}(r^{\beta-1}) }.
\ee

Using (\ref{afg}) to take the trace we see that $R \sim g^{\mu \nu } T_{\mu\nu}  \sim {\cal O}(r^{\beta-3})$.  From (\ref{afg}) we also find $\sqrt{-g}=r^{1-\beta}+{\cal O}(r^{-\beta})$, so that the integrand in the bulk part of (\ref{S}) is of order $r^{-2}$. It follows that the Einstein-Hilbert term is finite when evaluated on any smooth asymptotically flat 2+1 solution.

We now consider the boundary term on the timelike boundary $\partial {\cal M}_\infty$:
\be
{1 \over 8\pi G}\int_{\partial \cal{M}_\infty}\sqrt{-h}K. \label{bS}
\ee
By direct calculation for metrics with the asymptotic behavior (\ref{afg}) one finds
\be
\label{Kmat}
K_{\mu\nu} = \mat{K_{tt}& K_{t\theta} \\ K_{\theta t}& K_{\theta\theta} } \sim \mat{ {\cal O}(r^{\beta/2-2}) & {\cal O}(r^{-2}))
 \\ {\cal O}(r^{-2}) \ \ \ \ & \frac{1}{2} (2 -\beta) r^{1-\beta/2} },
\ee
so that to leading order in $r$ one has
\be
K=\frac{1}{2}(2-\beta)r^{-1+\beta/2}+{\cal O}(r^{-2+\beta/2}). \label{aK}
\ee
Using the fact that $\sqrt{-h}=r^{1-\beta/2}+{\cal O}(r^{1/2-\beta/2})$,
we see that the Gibbons-Hawking integrand is given by
\be
\sqrt{-h}K=1-\beta/2+ {\cal O}(r^{-1/2}). \label{aint}
\ee
It follows that, apart from past and future boundary terms,  the action (\ref{S}) is finite on asymptotically flat 2+1 solutions.

We now turn to stationarity of the action.  We use the fact that under a general variation one has
\be
\delta S_{HE+GH}={1 \over 16\pi G}\int_{\partial \cal{M}}\sqrt{-h}(K^{ij}-Kh^{ij})\delta h_{ij}. \label{dS}
\ee
Recall that $\delta h_{ij}$ vanishes on the past and future boundaries, while near $\partial {\cal M}_\infty$ we may
assemble the results (\ref{afh}), (\ref{invh}), and (\ref{Kmat}) to find 
\be
\sqrt{-h}(K^{ij}-Kh^{ij})\delta h_{ij} \sim  \sqrt{-h}(K^{tt}-Kh^{tt})\delta h_{tt}\sim {\cal O}(1/r), \label{adS}
\ee
which vanishes in the limit $r \rightarrow \infty$.  Thus, as claimed in section \ref{intro}, 
the action $S_{EH+GH}$ is stationary on 2+1 asymptotically flat solutions under arbitrary asymptotically flat variations.

\subsection{Comparison with higher dimensions}
\label{other}

We showed above that, for asymptotically flat spacetimes in 2+1 dimensions, the Gibbons-Hawking term provides a sufficient boundary term to promote the Einstein-Hilbert action to a satisfactory variational principle.  However, additional terms are needed in higher dimensions \cite{Mann:2005yr}.  Here we briefly discuss the extrapolation to 2+1 dimensions of  various additional boundary terms (called ``counter-terms'') which have been proposed for higher dimensional cases. We find that all such extrapolations either vanish identically or are ill-defined.  

The simplest asymptotically flat counter-term to discuss is the one proposed in \cite{Mann:1999pc}, proportional to $\int \sqrt{-h} \sqrt{\cal R}$, where ${\cal R}$ is the Ricci scalar of the boundary metric.  Because a 1+1 cylinder is flat, this term simply vanishes for 2+1 asymptotically flat geometries.  Similarly, one may consider the counter-term proposed in \cite{Kraus:1999di}, which is proportional to $\frac{ {\cal R}^{3/2} } {\sqrt{ {\cal R}^2 - {\cal R}^{ij} {\cal R}_{ij}}}$,
where ${\cal R}_{ij}$ is the Ricci tensor induced on $\partial {\cal M}_\infty$.   When this ratio is well-defined, it again vanishes in the limit in which the cut-off is removed.

A more complicated (but more general) counter-term was  introduced in \cite{Mann:2005yr}.  This term is given by 
\be
- \frac{1}{8 \pi G} \int_{\partial{\cal M}} \hat K, \label{khatterm}
\ee
with $\hat K$ defined implicitly by the equation
\be
{\cal R}_{ij}={\hat K}_{ij}{\hat K}-{\hat K}_{il}{\hat K}^l_j. \label{khat}
\ee
However, this equation
degenerates in 2+1 dimensions.  This is easiest to see by considering the stronger relation
\be
\label{GC1}
{\cal R}_{ikjl}={\hat K}_{ij}{\hat K}_{kl} -{\hat K}_{ik}{\hat K}_{jl}, \label{khat2}
\ee
which reduces to (\ref{khat}) when contracted with $h^{kl}$.  Note that the right hand side of (\ref{khat2}) has all of the symmetries of the Riemann tensor.  As a result, in 1+1 dimensions it has only a single independent component which in fact is equal to $\det(\hat K)$.  Thus, equation (\ref{khat}) can determine at most $\det (\hat{K})$ and not the desired trace of $\hat K_{ij}$.  We see that the counter-term (\ref{khatterm}) is not well-defined for 1+1 dimensional boundaries.

Finally, one may ask if we may use as a counter-term an additional Gibbons-Hawking term 
\be
- \frac{1}{8 \pi G} \int_{\partial{\cal M}} K_{Ref}, \label{refK}
\ee
evaluated on some reference background (see e.g., \cite{Brown:1992br, Brown:1994gs, Hawking:1995fd}).  This typically requires embedding the boundary $(\partial {\cal M}, h)$ in the background and using this embedding to define the reference extrinsic curvature $K_{Ref}$.  As described in \cite{Mann:2005yr}, such an approach fails in higher dimensions because a generic boundary cannot be isometrically embedded in a given reference background.  Roughly speaking, co-dimension one embeddings into a fixed manifold are specified by a single relation between the coordinates of the target manifold, while more than one function of the embedded manifold is required to specify a generic (fully gauge fixed) metric. 
However, in 1+1 dimensions we may always choose, e.g., conformal gauge, in which an arbitrary geometry {\it is} specified by a single free function.  Thus, this counting argument does not rule out the use of background subtraction for 2+1 dimensional bulk spacetimes.

Nonetheless, it remains far from clear that background subtraction can succeed in the 2+1 context.  The problem is that, at least locally,  the intrinsic geometry of $\partial {\cal M}$ does not determine the extrinsic curvature $(K_{ref})_{ij}$.  Indeed, we argued above that $(K_{ref})_{ij}$ cannot be determined from (\ref{GC1}).  Of course, an embedding would also satisfy the remaining Gauss-Codazzi equations 
\be
\label{GC2}
r_\mu R^{\mu j}_{\;\; kl}=D_{ k}{K_{ref}}^j_{\;l}-D_{l}{K_{ref}}^j_{\;k}, \label{gc2}
\ee
where $D_{k}$ is the covariant derivative compatible with $h_{ij}$ and $r^\mu$ is the unit vector normal to $\partial {\cal M}_\infty$.   However, since these are differential equations and our $\partial {\cal M}_\infty$ is a manifold with boundary, we still expect constants of integration in any solution to (\ref{GC2}).  Without some prescription for the dependence of such integration constants on the cut-off $\Omega$, the $\Omega \rightarrow \infty$ limit of the boundary term (\ref{refK}) is ill-defined.
Thus we see that any proposed counter-term from higher dimensional asymptotically flat space which is well-defined in 2+1 dimensions vanishes identically.

\section{The gravitational Energy}
\label{Hamiltonian}

We  have seen that the action (\ref{S}) is finite up to past and future boundary terms, and that an arbitrary asymptotically flat  variation of (\ref{S}) vanishes on the space of asymptotically flat  solutions.  We now compute the gravitational hamiltonian as the Legendre transform of the associated Lagrangian.  We proceed by performing a 2+1 deomposition of the action following \cite{Hawking:1995fd}, which in turn follows \cite{ADM1,ADM2,ADM3, Wald:1984rg} in addressing the bulk terms.   Finally, we quickly show that the same energy is obtained by considering the boundary stress tensor associated with the action (\ref{S}).

\subsection{The 2+1 decomposition and the gravitational Hamiltonian}

Let us begin by foliating the space-time with a family of spacelike surfaces $\Sigma_t$ labeled by the coordinate $t$. We introduce a vector $t^\mu$ such that $t^\mu \nabla_\mu t=1$. In terms of the unit normal vector to the surfaces $\Sigma_t$ we can decompose $t^\mu$ into the usual lapse function $N$ and shift vector $N^\mu$ through $t^\mu=Nn^\mu+N^\mu$.   We choose $\Sigma_t$ orthogonal to the 1+1 part of the asymptotic boundary\footnote{In fact, we choose the spatial regulator $\Omega$ such that, for sufficiently large $\Omega$,  $\Sigma_t$ is orthogonal to {\it each} regulated boundary $\partial {\cal M}_\Omega$.} $\partial {\cal M}_\infty$.  We assume there are no inner boundaries, so the total boundary consists of $\partial {\cal M}_\infty$ (tangent to $n^\mu$) together with initial and final surfaces for which $n^\mu$ is a unit normal. Below, we include in (\ref{S}) the Gibbons-Hawking terms on the initial and final surfaces as well as that on $\partial {\cal M}_\infty$.

To perform the 2+1 decomposition of the action, we express the three dimensional scalar curvature in terms of the induced scalar curvature ${\cal R}$ and the extrinsic curvature $K^t_{ij}$ of $\Sigma_t$. As a first step, we write
\be
R=2(G_{\mu\nu}-R_{\mu\nu})n^\mu n^\nu . \label{eeq}
\ee
{}From the usual initial value constraint, the first term is
\be
2G_{\mu\nu}n^\mu n^\nu={\cal R}-K^t_{\mu\nu}(K^{t})^{\mu\nu}+(K^{t})^2. \label{ic}
\ee
For the second term we use the identity
\be
R_{\mu\nu}n^\mu n^\nu=K^{t\; 2}-K^t_{\mu\nu}K^{t\;\mu\nu}-\nabla_\mu(n^\mu\nabla_\nu n^\nu)+\nabla_\nu(n^\mu\nabla_\mu n^\nu). \label{cdn}
\ee
When integrated over spacetime, the two total derivatives in (\ref{cdn}) yield boundary terms. The first is proportional to $n^\mu$, so it contributes only  on the initial and final surfaces. On these surfaces, it completely cancels the the Gibbons-Hawking term. The second term is orthogonal to $n^\mu$ so it will contribute only on $\partial {\cal M}_\infty$. Adding the Gibbons-Hawking term to the boundary terms (\ref{cdn}) and expressing the result in terms of the unit vector $r^\mu$ normal to $\partial {\cal M}_\infty$ one finds
\be
{1 \over 8\pi G}\int_{\partial {\cal M}_\infty }\sqrt{-h}(\nabla_\mu r^\mu - r_\nu n^\mu\nabla_\mu n^\nu)={1 \over 8\pi G}\int_{\partial {\cal M}_\infty}\sqrt{-h}(g^{\mu\nu} +n^\mu n^\nu)\nabla_\mu r_\nu, \label{tbi}
\ee
where $g_{\mu\nu}$ is the metric corresponding to the line element (\ref{afg}). Recognizing $q^{\mu \nu} =  g^{\mu\nu} + n^\mu n^\nu$ as the projector onto the surface $\Sigma_t$, we see that the integrand in (\ref{tbi}) is the trace of the extrinsic curvature of the curve $C_t=\Sigma_t \cap \partial \cal{M}$, in the surface $\Sigma_t$. We will call this trace $k$.

Thus, the action (\ref{S}) takes the form
\be
S=\int Ndt \lbrack {1 \over 16\pi G}\int_{\Sigma_t}\sqrt{q}({\cal R}+(K^{t})^2-K^t_{\mu\nu} (K^t)^{\mu\nu})+{1 \over 8\pi G}\int_{C_t}\sqrt{q_C} k \rbrack, \label{stS}
\ee
where $q$ is the determinant of the metric induced on $\Sigma_t$ and $q_C$ is the determinant of the metric induced on $C_t$ .

Recalling that the extrinsic curvature $K^t_{\mu\nu}$ of $\Sigma_t$ is given by
\be
K^t_{\mu\nu}={1 \over 2N}[\dot q_{\mu\nu}-2D_{(\mu}N_{\nu)}], \label{Kg}
\ee
where $D$ is the covariant derivative on $\Sigma_t$ compatible with $q_{\mu\nu}$, and the momentum $\tilde \pi^{\mu \nu}$ conjugate to $q_{\mu \nu}$ is, as usual, 
\be
\widetilde{\pi}^{\mu\nu}={\delta {\cal L} \over \delta \dot q_{\mu\nu}}={\sqrt{q} \over 16\pi G}(K^{t\;\mu\nu}-K^tq^{\mu\nu}). \label{pi}
\ee
Defining a un-densitized momentum $\pi^{\mu\nu}={16\pi G \over \sqrt{q}N}\widetilde{\pi}^{\mu\nu}$, we can write the action as
\bea
S &=&\int Ndt \lbrack {1 \over 16\pi G}\int_{\Sigma_t}\sqrt{q}(\pi^{\mu\nu}\dot q_{\mu\nu} + {\cal R}+\pi^2
-\pi_{\mu\nu}\pi^{\mu\nu}+2N_\mu D_\nu\pi^{\mu\nu}) +{1 \over 8\pi G}\int_{C_t}\sqrt{q_{C}} k \rbrack \cr &-&{1 \over 8\pi G} \int dt \int_{C_t} \sqrt{q_{C}} N^\mu r^\nu \pi_{\mu\nu} . \label{stSpi}
\eea
It is now easy to Legendre transform the Lagrangian to obtain
\bea
H &=&-{1 \over 16\pi G}\int_{\Sigma_t}N\sqrt{q}({\cal R}+\pi^2
-\pi_{\mu\nu}\pi^{\mu\nu}+2N_\mu D_\nu\pi^{\mu\nu}) \cr &-&{1 \over 8\pi G}\int_{C_t} (N\sqrt{q_{C}}k -\sqrt{q_C}N^\mu r^\nu \pi_{\mu\nu}). \label{H}
\eea

However, the boundary term simplifies further when we make explicit use of the asymptotic behavior (\ref{afg}). The first term of the boundary integrand is, to leading order, the same as we computed in (\ref{aint}). For the second term we have
\be
\pi_{ij} \sim \mat{\pi_{tt} & \pi_{t\theta} \\ \pi_{\theta t} & \pi_{\theta \theta} }\sim 
\mat{ {\cal O}(r^{-1-\beta})  &  \ {\cal O}(r^{-1-\beta}) \\  {\cal O}(r^{-1-\beta}) & {\cal O}(r^{1-\beta})}, \label{api}
\ee
while the vectors $N^\mu$ and $r^\mu$ satisfy
\be
N^\mu\sim{\cal O}(r^{-1-\beta}),\;\;\; r^r=r^{\beta/2}+{\cal O}(r^{-1-\beta})\;\;\;\;\; {\rm and} \;\;\;\;\; r^\theta\sim{\cal O}(r^{-1-\beta}). \label{Nr}
\ee
Thus the second term in the integral over $C_t$ is of order ${\cal O}(r^{-4\beta})$ and vanishes as $r \rightarrow \infty$.

As a result, we obtain
\be
H=-{1 \over 16\pi G}\int_{\Sigma_t}N\sqrt{q}({\cal R}+\pi^2
-\pi_{\mu\nu}\pi^{\mu\nu}+2N_\mu D_\nu\pi^{\mu\nu})-{1 \over 16\pi G}\int_{C_t}(2-\beta) . \label{aH}
\ee
This coincides with the Hamiltonian obtained in  \cite{Ashtekar:1993ds} via Regge-Teitelboim methods \cite{RT}, except for the addition of the constant term $- \frac{1}{16\pi G} \int_{C_t} 2 = - \frac{1}{4G}.$

\subsection{Energy from the boundary stress tensor}

By using the Hamiltonain form of the action (\ref{stSpi}), it is straightforward to compare the
ADM expressions   (\ref{aH}) for the energy with that defined via
a boundary stress tensor.  Recall 
\cite{DAR,Mann:2005yr} that while for asymptotically flat spacetimes one cannot vary the on-shell action with respect to boundary conditions, a useful boundary stress tensor can nevertheless be defined by first considering the action $S_\Omega$ of each regulated spacetime ${\cal M}_\Omega$ defined in section
\ref{action}.  One defines
\begin{equation}
T_{\mu \nu} (\Omega) = - \frac{2}{\sqrt{-h}} \frac{\delta S_\Omega}{\delta h^{\mu \nu}} = -
\frac{2}{\sqrt{-h}} \frac{\delta S_\Omega}{\delta g^{\mu \nu}}, 
\end{equation}
and the boundary stress tensor energy is defined by  
\be
E_{bst} = \lim_{\Omega \rightarrow \infty} \int_{C_t} \sqrt{q_C}\  n_\mu n_\nu T^{\mu\nu}. 
\ee

To compare with the Hamiltonian definition of Energy (\ref{aH}), we simply express $T_{\mu \nu}$ in terms of variations of the action with respect to the lapse
$N$ and shift $N^\mu$, with the spatial metric $q_{\mu \nu} = g_{\mu \nu} + n_\mu
n_\nu$ held constant.  To this end, it is useful to compute certain
partial derivatives.  Using the relation $dt = - \frac{1}{N} n_\mu dx^\mu$
we find
\begin{equation}
\frac{\delta g_{\mu \nu}}{\delta N} \Bigg |_{q_{\kappa \lambda}}  =
\frac{\delta}{\delta N} \left( - n_\mu n_\nu \right)  =
\frac{-2}{N} n_\mu n_\nu.
\end{equation}
As a result, we have
\begin{eqnarray}
E_{bst} &=& \int_{C_t} \sqrt{q_C} \  n_\mu n_\nu T^{\mu\nu} = \int_{C_t}
\sqrt{q_C} \ \frac{2n_\mu n_\nu}{ \sqrt{-h }} \frac{\delta S}{\delta
g_{\mu \nu}} \cr &=&  - \int_{C_t} \frac{\delta g_{\mu \nu}}{\delta N}
\frac{\delta S}{\delta g_{\mu \nu}} = - \int_{C_t}  \frac{\delta
S}{\delta N} \Bigg |_{q_{\mu \nu}}  = - \frac{1}{16 \pi G} \int_{C_t} (2-\beta).
\end{eqnarray}
Since the bulk contribution to (\ref{aH}) vanishes on the constraint surface, it is clear that the Hamiltonian and boundary counter-term definitions of energy agree\footnote{Note that, assuming that the appropriate actions can be written in an appropriate canonical form, this argument provides a simple derivation of the main result of \cite{HIM2} and of certain results from \cite{HIM1,PS}}.

\section{Discussions}
\label{disc}

We have shown above that the Einstein-Hilbert action with Gibbons Hawking term (\ref{S}) provides a satisfactory variational principle for 2+1 asymptotically flat spacetimes.  In higher dimensions it is important to add more complicated boundary terms but, as discussed in section \ref{other}, these do not extend naturally to 2+1 dimensions. Defining the boundary by a cylindrical cut-off prescription, we have shown that on the space of the solutions the action i) is finite (apart from the past and future boundary terms) and ii) is stationary under any asymptotically flat variation.

Because (\ref{S}) is appropriately finite and defines a good variational principle, a Hamiltonian can be defined directly via the Legendre transform of the Lagrangian.
The result of this Legendre transform is (\ref{aH}), which agrees with the standard result \cite{Ashtekar:1993ds} except for a shift of the zero of energy.  Interestingly, this shift causes our Hamiltonian to take values in the range
\begin{equation}
H \in [-\frac{1}{4 G}, 0],
\end{equation}
and in particular, assigns 2+1 Minkowski space the energy $E_{Mink_{2+1}} =  -\frac{1}{4 G}$.  
Such a non-zero energy is possible because the  phase space of asymptotically flat gravity in 2+1 dimensions is not invariant under Lorentz transformations \cite{Ashtekar:1993ds,MH}.

As noted above, our shift sets the upper bound on the energy of 2+1 asymptotically flat spacetimes to zero.  Physically, the upper bound arises because, at this value of the energy, the deficit angle at infinity reaches $2 \pi$ and the asymptotic region ``closes off''.  Now, recall that any spatially closed universe is naturally assigned zero energy as well.  In particular, since there is no boundary ${\cal M}_\infty$, this assignment follows from a definition of the Hamiltonian for such systems via a Legendre transform of the Lagrangian as above; one finds that the Hamiltonian is constrained to vanish.  This continuity property of the energy when interpolating between spatially compact and asymptotically flat spacetimes may merit further investigation. 

To readers familiar with asymptotically AdS counter-terms (e.g., \cite{skenderis,Balasubramanian:1999re}), it may not seem surprising that (\ref{S}) provides a valid variational principle.  In 2+1 dimensions, the only additional AdS counter-term required (beyond the terms in (\ref{S})) is proportional to $\frac{1}{\ell} \int_{\partial {\cal M}} \sqrt{-h}$, where $\ell$ is the AdS length scale.  Thus, this additional AdS counter-term vanishes as $\ell \rightarrow \infty$.  In contrast, in higher dimensions AdS boundary conditions require boundary terms proportional to positive powers of $\ell$, and so do not readily admit a flat space limit.

Similarly, to AdS-familiar readers, it may not be surprising that 2+1 Minkowski space has a negative vacuum energy.  In 2+1 dimensions, adding the appropriate covariant counter-terms and using the boundary stress tensor prescription \cite{Balasubramanian:1999re} yields an energy $E_{AdS_{2+1}} = -1/8G$ for 2+1 AdS space, and in particular one  which is independent of the AdS length scale $\ell$.  (In contrast, in higher dimensions the energy of AdS space vanishes as $\ell \rightarrow \infty$.)    However, what {\it is} quite striking is that the energy $E_{Mink_{2+1}} = -1/4G$ of 2+1 Minkowski space is {\it not} the limit as $\ell \rightarrow \infty$ of the AdS result.  Instead, it differs by a factor of 2.   One may hope that a deeper understanding of this result is related to a deeper understanding of the flat space limit of AdS/CFT.

One may ask how unique is our prescription for calculating the energy.  After all, 
we have obtained the result (\ref{aH}) from a particular choice of boundary terms in the covariant action (\ref{S}).  Now, for asymptotically AdS spaces in 2+1 dimensions, 
 the boundary terms are {\it entirely} determined by the properties of locality and covariance \cite{Balasubramanian:1999re}.  Thus the result
$E_{AdS_{2+1}} = -1/8G$ does not suffer from the ``scheme dependence'' that one sees in higher dimensions.  We will now show that our results are similarly unique given the requirement that one begin with a well-defined variational principle built from a local covariant boundary term.

To this end, 
consider any choice of boundary term built
locally from the extrinsic curvature $K_{ij}$ and the Riemann curvature ${\cal R}_{ijk}{}^l$ of $\partial {\cal M}_\infty$
and which leads to a valid variational principle for asymptotically flat spacetimes.  Any such boundary term
will differ from the one in (\ref{S}) by some finite $\Delta S$ whose variation vanishes within the class 
of asymptotically flat spacetimes. In particular, it must be invariant under (time-dependent) variations of the deficit angle at infinity,
\begin{equation}
\label{varybeta}
\beta \rightarrow \beta + \delta \beta(t).
\end{equation}
To see the effect of this requirement, let us evaluate $\Delta S$ on the point particle spacetimes (\ref{pp})
using $r=constant$ surfaces to compute the boundary term.  The Riemann tensor vanishes on such surfaces, and the only
non-zero component of $K_{ij}$ is $K_{\theta \theta} = \frac{1}{2} (2-\beta) r^{1-\beta/2}$.  The boundary metric itself can enter only through $\sqrt{-h}$ and $h^{\theta \theta}$, and so provides only factors of $ r^{\pm(1-\beta/2)}$. If $\Delta S$ is to have a well-defined limit at large $r$, the $r$-dependence of the metric can serve only to cancel the $r$-dependence of $K_{\theta \theta}$.  
As a result, we have
$\Delta S = f(K_{\theta \theta} r^{-(1-\beta/2)}) (T_+ - T_-) = f(2-\beta) (T_+ - T_-)$.   

Now, consider a variation $\delta \beta(t)$ which is constant over a time interval $\Delta t$ constituting most of interval $[T_-,T_+]$, but which vanishes at $t= T_\pm$ to preserve the past and future boundary conditions.  Since the variation of $\Delta S$ under (\ref{varybeta}) contains a term proportional to $ \frac{df}{d\beta} \Delta t$, our $\Delta S$ can be stationary  only if $f= constant$; i.e, 
if on such solutions $f$ is independent of $K_{ij}$.  But since the Riemann tensor on $\partial {\cal M}$ vanishes, 
on such solutions we must have
\begin{equation}
\label{bndyvol}
\Delta S = c \int_{\partial {\cal M}_\infty} \sqrt{-h}.
\end{equation}
Imposing either the requirement that $\Delta S$ be finite or that $\Delta S$ be stationary then sets $c=0$, and thus $\Delta S=0$ on such solutions.
But since $\Delta S$ is stationary under all asymptotically flat variations, it is in fact constant on the space of 
solutions.  Thus, $\Delta S=0$ on all solutions.  As a result, it cannot affect the result (\ref{aH}) for the
energy.

On dimensional grounds, a vacuum energy such as we have found can arise only in 2+1 dimensions, where $c^2/G$ has dimensions of mass (without using any factors of $\hbar$).  Indeed, calculations of the vacuum energy of $d+1$ Minkowski space from a covariant variational principle for $d>2$ obtain $E_{Minkowski}=0$  \cite{DAR,Mann:2005yr}.  Here comparison with AdS space is more subtle, as the AdS vacuum energy depends  \cite{Balasubramanian:1999re} on the particular choice of boundary term (i.e., the choice of ``renormalization scheme") used to define the covariant variational principle. Nevertheless, it is interesting to note that in the most common scheme the vacuum energy diverges in the limit $\Lambda \rightarrow 0$.  This again suggests subtleties in the flat space limit of AdS/CFT which would be interesting to understand in detail.

\section{Acknowledgments}

The authors wish to thank Jim Isenberg and Greg Galloway for input on the two-dimensional embedding problem, and Bernd Schroers for other useful discussions.
The authors were supported in part by NSF grants
PHY0354978 and PHY99-07949, and in part by funds from the University of California. 
 L.P. was also supported by the Postdoctoral Research Fellowship number FE-05-45 provided by UC Mexus-Conacyt. D. M. wishes to thank the Kavli Insititute of Theoretical Physics for its hospitality during this work.

\end{document}